\documentclass[runningheads]{llncs}
\usepackage[T1]{fontenc}
\usepackage{graphicx}
\usepackage{bm}
\usepackage{hyperref}

\graphicspath{{./img}{./}}

\usepackage{amsmath}
\usepackage{color}

\begin{document}
\title{Approximating robot reachable space using convex polytopes}
%
%
\author{Antun Skuric \and
Vincent Padois\and
David Daney}
\authorrunning{A. Skuric et al.}
%
\institute{INRIA Sud-Ouest, Bordeaux, France \\
\email{firstname.lastname@inria.fr}\\
\url{http://team.inria.fr/auctus/} \\
This work was funded by the BPI France Lichie project.}
\maketitle              
\begin{abstract}

This paper presents an approach for approximating the reachable space of robotic manipulators based on convex polytopes. The proposed approach predicts the reachable space over a given time horizon based on the robot's actuation limits and kinematic constraints. The approach is furthermore extended to integrate the robot's environment, assuming it can be expressed in a form of linear constraints, and to account for the robot's link geometry.

The accuracy of the proposed method is evaluated using simulations of robot's nonlinear dynamics and it is compared against the cartesian space limits, usually provided by manufacturers in standard datasheets.

The accuracy analysis results show that the proposed method has good performance for the time horizons up to $t_h\!\leq$250ms, encapsulating most of the simulated robot's reachable space while maintaining comparable volume. For a 7 dof robot, the method has an average execution time of 50ms, independent of the horizon time, potentially enabling real-time applications.


\end{abstract}
\section{Introduction}
\vspace{-0.3cm}

Collaborative robots, designed for safe physical interaction with humans, have a great potential to find their way in many spheres of industry, research and potentially become a part of our everyday lives. Apart from replacing rigid and dangerous manipulators in the industrial applications, their main potential is creating new robotic assistance scenarios leveraging high degree of physical human-robot interaction. Such scenarios demand not only many safety guarantees, but a high degree of operator's situation awareness both in terms of robot's behaviour and its physical abilities in real-time.  

A metric unifying the robot's physical abilities and its movement capacity, important for safety and performance analysis, is called the robot's \textit{reachable space} \cite{althoff2014}. It represents the space of reachable robot's positions (ex. end-effector cartesian positions) for a certain time horizon, given different assumptions on its physical abilities. As robotic systems are highly non-linear and featuring complex dynamics, the true reachable space of robots is very complex to characterise and calculate. Therefore, approximation techniques are necessary in order to yield practical solutions. Pereira and Athhoff\cite{pereira2017} have developed an interval analysis based approach, later improved by Shrepp et al.\cite{schepp2022}, to approximate the human arm reachable space, modeled as a serial robotic manipulator, using a set of spheres and cylinders. However, in many cases the approximation of the reachable space using these shapes is impractical due to their non-linear nature.

In this paper, an approximation approach of the robot's reachable space using convex polytopes is proposed, leveraging several of their key characteristics. Convex polytopes can be represented as a set of linear inequalities $A\bm{x}\leq\bm{b}$, which may be directly used with different optimisation techniques. Additionally, this set of linear inequalities can be intuitively extended with different environmental and user defined constraints. Since the reachable space is often low dimensional ($\leq$3D) these polytopes can be easily visualised, having the structure of a triangulated mesh, and potentially provide a human operator with valuable information about the robot's capacity. Furthermore, operations over polytopes such as Minkowski sum and intersection are well defined and efficient to calculate, enabling for the intuitive extension of the proposed method to account for robot's link geometry. Finally, the efficiency of polytope enumeration techniques for low dimensional polytopes has a potential to make this metric real-time capable. 

Even though convex polytopes are widely used to characterise different metrics of robot's physical abilities\cite{skuric2021}: force, acceleration  and velocity, to the best of our knowledge they have not yet been used to characterise the reachable space of a robotic manipulator. 
The proposed approach is partially inspired by the works of Long et al. \cite{long2018} on the constrained manipulability (velocity) polytopes.  

Depending on whether the application requires an under or over approximation, or perhaps time-efficiency, the evaluation metrics of the accuracy of the reachable space approximation will differ significantly. In this paper, in addition to the execution time analysis, three accuracy measures are chosen to enable the evaluation of the proposed approach's performance and study its limitations.

The paper has the following structure. In the chapter \ref{ch:polytope} the formal definition of the approach is introduced, followed by the chapter \ref{ch:analysis}, describing the procedure of benchmarking and the accuracy analysis. In the chapter \ref{ch:results}, the results of the accuracy and execution time analysis of the approach are given and the chapter \ref{ch:discussion} discusses the main limitation as well as potential applications.


\vspace{-0.1cm}
\section{Convex polytope of reachable space}
\label{ch:polytope}
\vspace{-0.2cm}

Robot dynamics in joint space can be expressed as
\begin{equation}
    M(\bm{q})\ddot{\bm{q}} +\underbrace{ C(\bm{q},\dot{\bm{q}})\dot{\bm{q}} + \bm{\tau}_g(\bm{q}) }_{\bm{\tau}_d}= \bm{\tau}
    \label{eq:robot_model}
\end{equation}
where $M$ is the mass matrix, $C$ is coriolis matrix and $\bm{\tau}_g$ is the gravity vector. $\bm{\tau}$ is the applied joint torque vector and $\bm{\tau}_d$ is the equivalent joint torque vector due to the coriolis and gravity effects.

For a robot with $n$ actuated degrees of freedom (DOF), the robot's joint torque $\bm{\tau}$, velocity $\dot{\bm{q}}$ and joint angles $\bm{q}$ are $n$-dimensional vectors, limited by the robot's hardware\footnote{There is a coupling between the velocity and torque limits of an actuator related to its power. Yet, it is common practice to consider a subset of these limits such as torque and velocity can be chosen independantly one from the other.}
\begin{equation}
 \bm{\tau}\in\left[\bm{\tau}_{min},\bm{\tau}_{max}\right], \quad 
 \dot{\bm{q}}\in\left[\dot{\bm{q}}_{min},\dot{\bm{q}}_{max}\right], \quad \bm{q}\in\left[\bm{q}_{min},\bm{q}_{max}\right]
 \label{eq:limits}
\end{equation}

For a given moment in time $k$ and for given robot configuration $\bm{q}_k$, the affine relationship between joint torques $\bm{\tau}$ and accelerations $\ddot{\bm{q}}_{k}$ can be expressed as
\begin{equation}
    \ddot{\bm{q}}_{k} = M^{-1}(\bm{q}_k)(\bm{\tau} - \bm{\tau}_d) = M_k^{-1}(\bm{\tau} - \bm{\tau}_d)
\end{equation}

Considering fixed joint acceleration $\ddot{\bm{q}}_{k}$ during given horizon length $t_h$, an approximation of the robot's joint velocity $\dot{\bm{q}}_{k+1}$ and position $\bm{q}_{k+1}$, at the end of horizon, can be calculated using numerical integration (forward Euler method) 
\begin{equation}
    \dot{\bm{q}}_{k+1} = M_k^{-1}t_h(\bm{\tau} - \bm{\tau}_d) + \dot{\bm{q}}_{k}, \qquad \bm{q}_{k+1} = M_k^{-1}\frac{t_h^2}{2}(\bm{\tau} - \bm{\tau}_d) + \dot{\bm{q}}_{k}t_h + \bm{q}_{k}
    \label{eq:joint_vel_pos}
\end{equation}
This linear numerical integration (\ref{eq:joint_vel_pos}) considers the robot's dynamics (\ref{eq:robot_model}), and the applied joint torque $\bm{\tau}$, fixed during the horizon time $t_h$. Such assumption is reasonable only for short horizon times $t_h$ which in term represents a limitation of the proposed method.

The relationship between the $m$-dimensional task space velocity and acceleration of certain frame on the robot (for example end-effector frame) and the $n$-dimensional joint space equivalents is defined through the corresponding jacobian matrix $J(\bm{q})$ and its time derivative $\dot{J}(\bm{q})$. 
\begin{equation}
    \dot{\bm{x}}_k = J(\bm{q}_k)\dot{\bm{q}}_k = J_k \dot{\bm{q}}_k, \qquad  \ddot{\bm{x}}_k = J_k \ddot{\bm{q}}_k + \dot{J}_k \dot{\bm{q}}_k
\end{equation}

Finally, given the horizon of interest $t_h$, the predicted cartesian position $\bm{x}_{k+1}$ can then be expressed as

\begin{equation}
\begin{split}
    {\bm{x}}_{k+1} &= \ddot{\bm{x}}_k\frac{t_h^2}{2}   \qquad\qquad\qquad\qquad\qquad\qquad\qquad\qquad+\dot{\bm{x}}_kt_h \quad +\!  \bm{x}_k\\
    &=  J_k M_k^{-1}\frac{t_h^2}{2}\bm{\tau} \quad \underbrace{\underbrace{-
    J_k M_k^{-1}\frac{t_h^2}{2}\bm{\tau}_d}_{\Delta \bm{x}_{k,dyn}}  \quad \underbrace{ +\dot{J}_k \dot{\bm{q}}_k\frac{t_h^2}{2} \quad+ \dot{\bm{x}}_kt_h}_{\Delta \bm{x}_{k,vel}} \quad + \bm{x}_{k} }_{\bm{x}^*_{k+1}}
    \end{split}
    \label{eq:pred_pos}
\end{equation}

Where the $\bm{x}^*_{k+1}\! =\!\Delta \bm{x}_{k,dyn}\! +\! \Delta \bm{x}_{k,vel} + \bm{x}_{k}$ is a predicted position vector based on the joint configuration $\bm{q}_k$ , joint velocity $\dot{\bm{q}}_k$ (or cartesian velocity  $\dot{\bm{x}}_k\!=\!J\dot{\bm{q}}_k$) and the joint torque $\bm{\tau}_d$ in the current time-step $k$, while the first term describes the influence of the applied joint torques $\bm{\tau}$.

The convex polytope of reachable cartesian space $\mathcal{P}_x$ can then be defined as a set of all the possible cartesian positions $\bm{x}_{k+1}$ at the end of the horizon $t_h$, achieved by any combination of joint torques $\bm{\tau}$ within robot's physical limitations (\ref{eq:limits}), given the robot's current state $\dot{\bm{q}}_k$, $\bm{q}_k$ and $\bm{\tau}_d$ .

\begin{figure}[!t]
    \centering
    \includegraphics[width=\linewidth]{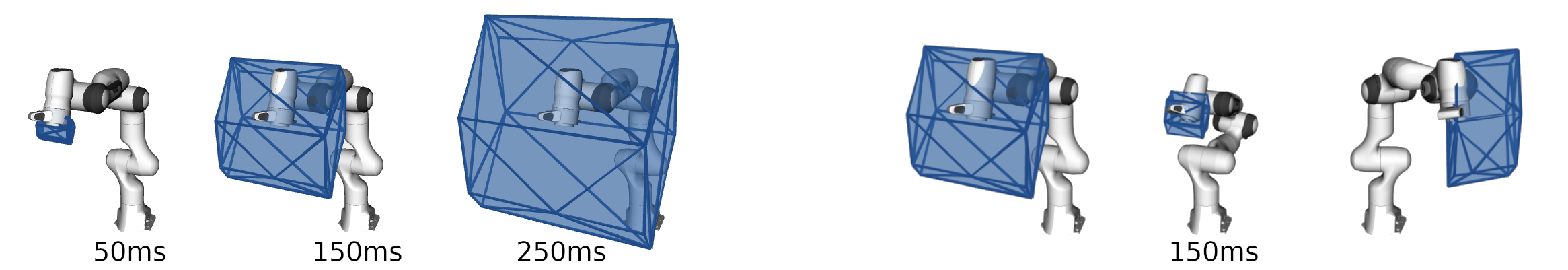}
    \caption{Three images on the left show the comparison of the size of the reachable space polytope $\mathcal{P}_x$ of a \textit{Franka Emika Panda}\protect\footnotemark robot's end-effector for 3 horizons  (left to right) $t_h$ = 0.05, 0.15 and 0.25s, for the same configuration. Whereas the three figures on the right show the constraining effect of the joint position limits on the $\mathcal{P}_x$ for a fixed horizon time ($t_h$=0.15s).
    Third image from the right shows the robot in its initial position $\bm{q}=[0,0,0,-\pi/2, 0, 3\pi/5,0]$, close to the center of the joint ranges. In the second image form the right, the robot's joints $q_1$ and $q_3$ at their limits, $\bm{q}=[0,-1.59,0,-2.9,0,3\pi/5,0,0]$. On the last image, robot's joints $q_0$ and $q_2$ are at their limits $\bm{q}=[-2.72,0, -2.72,-\pi/2,0,13\pi/5,0,0]$, preventing the robot to rotate around the z axis in one direction.} 
    \label{fig:horizon}
\end{figure}
\footnotetext{More information about the Panda robot at \url{https://frankaemika.github.io/docs/}}


\begin{figure}[!t]
    \centering
    \includegraphics[width=\linewidth]{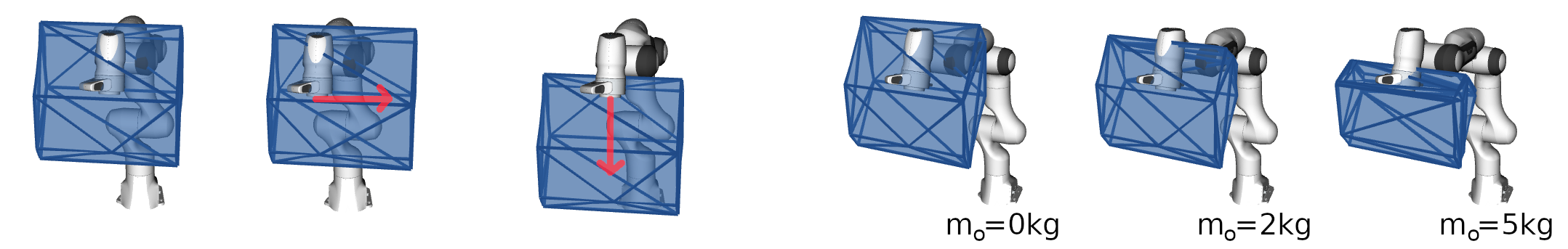}
    \caption{Three images on the left show the shifting effect on the reachable space polytope $\mathcal{P}_x$ produced by certain cartesian velocity at the beginning of the horizon, from left to right: $\dot{\bm{x}}_k$ = $[0,0,0]$, $[0,1,0] $ and $[0,0,-1]m/s$. Three images on the right show the reducing effect on the polytope $\mathcal{P}_x$ by three different carried object masses (left to right) $m_o$ = 0, 2 and 5kg. Horizon used is $t_h$=0.15s and the robot is in initial configuration.}
    \label{fig:velocity}
    \vspace{-1cm}
\end{figure}


\begin{equation}
\begin{split}
    \mathcal{P}_x = \{ \bm{x}_{k+1} \in \mathbf{R}^m \quad| \quad \bm{x}_{k+1} &= J_k M_k^{-1}\frac{t_h^2}{2}\bm{\tau} + \bm{x}^*_{k+1},\\
    \quad \bm{\tau} &\in \left[\bm{\tau}_{min},\bm{\tau}_{max}\right],\\
   M_k^{-1}t_h (\bm{\tau} - \bm{\tau}_d) + \dot{\bm{q}}_{k} &\in \left[\dot{\bm{q}}_{min},\dot{\bm{q}}_{max}\right],\\
   M_k^{-1}\frac{t_h^2}{2}(\bm{\tau} - \bm{\tau}_d) +  \dot{\bm{q}}_{k}t_h + \bm{q}_{k} &\in \left[\bm{q}_{min},\bm{q}_{max}\right] \}
\end{split} 
\label{eq:polytope_simple}
\end{equation}

Figures \ref{fig:horizon} and \ref{fig:velocity} showcase the influences of different horizon lengths $t_h$, robot constraints and robot movement ($\dot{\bm{q}}_k,\bm{\tau}_d$) on the shape of $\mathcal{P}_x$. Chapter \ref{ch:enumerating} describes an approach to enumerate the reachable space polytope.

\vspace{-0.3cm}
\subsection{Influence of the carried object}

If the robot is carrying a payload, an object with a mass $m_o$ and inertia $i_o$ attached to its end-effector, this object will have an influence on the robot's dynamics, modifying the mass matrix $M$, coriolis matrix $C$ and the gravity vector $\bm{\tau}_g$ \cite{hamad2019}.
The augmented dynamical model of the robot will have reduced acceleration capabilities due to the added effort necessary for the object's movements.

The figure \ref{fig:velocity} shows the influence of different object masses $m_o$, on the resulting reachable space polytope $\mathcal{P}_x$ for the \textit{Franka Emika Panda} robot. 






\vspace{-0.2cm}
\subsection{Integration of the environment}

\begin{figure}[!t]
    \centering
    \includegraphics[width=0.75\linewidth]{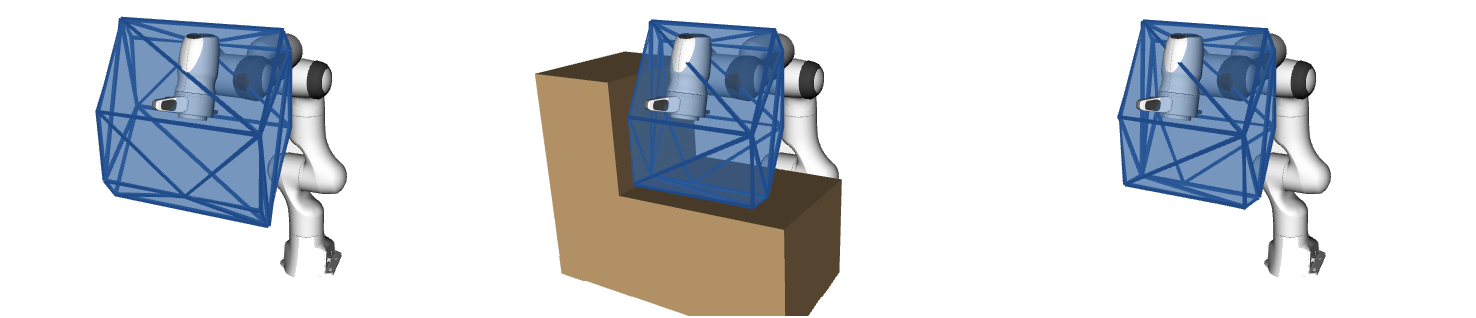}
    \caption{Resulting reachable space polytope $\mathcal{P}_x$ for a \textit{Franka Emika Panda} robot's end-effector when integrating the environmental constraints, horizon time used is $t_h$=0.15s. Environment is defined as $z\geq0.5$m and $y\geq-0.2$m.}
    \label{fig:env}
    \vspace{-0.5cm}
\end{figure}

If the robot's environment can be defined as a set of convex inequalities
\begin{equation}
    A_e \bm{x} \leq b_e
\end{equation}
these constraints can be directly transformed to the constraints of the joint torque $\bm{\tau}$ using the equation (\ref{eq:pred_pos})
\begin{equation}
    A_e\bm{x}_{k+1} = A_e J_k M_k^{-1}\frac{t_h^2}{2} (\bm{\tau} -\bm{\tau}_d) + A_e \bm{x}^*_{k+1} \leq b_e
    \label{eq:env_limits}
\end{equation}
and then included in the $\mathcal{P}_x$ calculation (\ref{eq:polytope_simple})
\begin{equation}
\begin{split}
    \mathcal{P}_x = \{ \bm{x}_{k+1} \in \mathbf{R}^m \quad| \quad \bm{x}_{k+1} &= J_k M_k^{-1}\frac{t_h^2}{2}\bm{\tau} + \bm{x}^*_{k+1},\\
   \quad \bm{\tau} &\in \left[\bm{\tau}_{min},\bm{\tau}_{max}\right],\\
    M_k^{-1}t_h (\bm{\tau} -\bm{\tau}_d)+ \dot{\bm{q}}_{k} &\in \left[\dot{\bm{q}}_{min},\dot{\bm{q}}_{max}\right],\\
   M_k^{-1}\frac{t_h^2}{2}(\bm{\tau} -\bm{\tau}_d) +  \dot{\bm{q}}_{k}t_h + \bm{q}_{k} &\in \left[\bm{q}_{min},\bm{q}_{max}\right],\\
   A_e J_k M_k^{-1}\frac{t_h^2}{2} (\bm{\tau} -\bm{\tau}_d) + A_e \bm{x}^*_{k+1} &\leq b_e ~~\}
\end{split} 
\label{eq:env}
\end{equation}

Figure \ref{fig:env} showcases the influence of the environmental constraints on the shape of the $\mathcal{P}_x$.

\vspace{-0.2cm}
\subsection{Integration of robot's link geometry}

If the robot link $l_i$ is considered to be a straight line, to calculate the predicted reachable space of a robot's link $l_i$ the polytope $\mathcal{P}_x$ can be calculated at the start $\mathcal{P}_{xs}$ and the end $\mathcal{P}_{xe}$ of the line $l_i$.  Finally the reachable space of this idealised robot link $l_i$ can then be calculated as the \textit{convex-hull} (CH) of the polytopes $\mathcal{P}_{xs}$ and $\mathcal{P}_{xe}$
\begin{equation}
    \mathcal{P}_{xli} = CH \left(  \mathcal{P}_{xs}, ~ \mathcal{P}_{xe} \right)
\end{equation}

If, instead of a straight line, a space captured by a robot's link $l_i$ can be expressed as a convex set of constraints or a polytope
\begin{equation}
    \mathcal{L}_i = \Big \{ \bm{x}_l ~ |~ A_l \bm{x}_l \leq b_l \Big\}
\end{equation}
the reachable space polytope $\mathcal{P}_{xli}$ can be extended to account for this link geometry by first finding the vertices of the polytope $\mathcal{L}$, followed by computing the reachable space polytope $\mathcal{P}_{x}$ for each one of the $k$ vertices, and finally computing their convex hull. 

\begin{equation}
    \mathcal{P}_{xl,_i} = CH \left(  \mathcal{P}_{x1}, ~ \dots ~, \mathcal{P}_{xk} \right)
\end{equation}

\begin{figure}[!t]
    \centering
    \includegraphics[width=\linewidth]{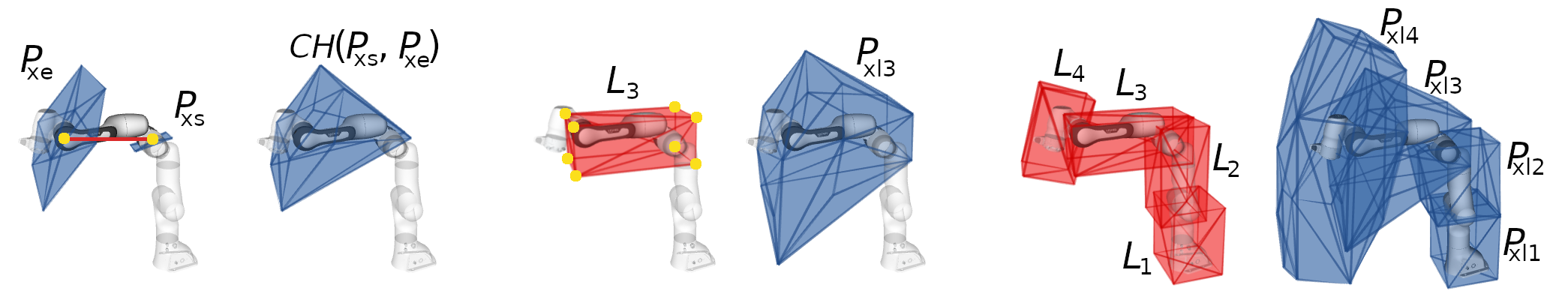}
    \caption{ First four images show the construction of the reachable space polytope of the panda robot's link 3, $\mathcal{P}_{xl3}$. From left to right they show, robot's link 3 as a line segment (in red) with polytopes $\mathcal{P}_{xs}$ and $\mathcal{P}_{xe}$, their convex-hull, an example of the enveloping space $\mathcal{L}_3$ of the link 3 (in red), and the final polytope $\mathcal{P}_{xl3}$ calculated as a convex hull of the reachable space polytopes of each one of the 8 vertices (shown in yellow) of $\mathcal{L}_3$.
    Last two images further show an example of 4 enveloping spaces $\mathcal{L}_i$ for each one of the robot's links, and their reachable space polytopes $\mathcal{P}_{xli}$. Robot is in its initial configuration, and the horizon time used is 150ms.}
    \label{fig:minkowski}
\vspace{-0.3cm}
\end{figure}

By evaluating the polytope $\mathcal{P}_{xli}$ for each robot's link one can obtain an efficient approximation of the envelope of the reachable space of the entire robot, as shown on figure \ref{fig:minkowski}.



\vspace{-0.2cm}
\subsection{Enumerating reachable space polytope}
\label{ch:enumerating}

The reachable space polytope $\mathcal{P}_x$ belongs to the family of over-determined problems
\begin{equation}
    \mathcal{P}_x = \{ \bm{x} \in \mathbf{R}^m \quad| \quad \bm{x}=P\bm{\tau},\quad  A\bm{\tau}\leq \bm{b} ~~\}, \quad \bm{\tau}\in R^n, \quad n\geq m
\label{eq:polytope_family}
\end{equation}
where the inequality constraint $A\bm{\tau}\leq \bm{b}$ encapsulates all the robot constraints (\ref{eq:limits}) as well as the environmental constraints (\ref{eq:env_limits}), and the matrix $J_k M_k^{-1}\frac{t_h^2}{2}$ becomes the projection matrix $P$. Vector $\bm{\tau}$ corresponds to the joint torque vector and the vector $\bm{x}$ corresponds to the difference $\bm{x}_{k+1}\!-\!\bm{x}_{k+1}^*$.

In order to fully define and simplify this formulation, the set of all the vertices ($\mathcal{V}$-representation) or all the half-spaces ($\mathcal{H}$-representation) of this polytope has to be found. Vertex representation of a polytope is a list of all of its vertices $\bm{v}_i$ and is commonly used for visualisation purposes
$
\mathcal{V}\!= \!\{ \bm{v}_0, \bm{v}_1, \bm{v}_2, \dots \}
$,
while half-space representation is defined as a list of inequalities $H\bm{x} \leq \bm{d}$, each corresponding to one of the polytope facets. This matrix inequality equation is more suitable for applications involving different optimisation strategies.

The reachable space polytope (\ref{eq:polytope_family}) is a particular case of the projection of a high-dimensional polytope onto a lower dimensional space. Therefore, the set of constraints $A\bm{\tau}\leq\bm{b}$ form a $\mathcal{H}$-representation of a polytope $\mathcal{P}_\tau$ in the $n$-dimensional joint torque space.
\begin{equation}
    \mathcal{P}_\tau = \{ \bm{\tau} \in \mathbf{R}^n \quad| \quad A\bm{\tau}\leq \bm{b} ~~\}
\label{eq:polytope_torque}
\end{equation}

This polytope is then projected using the projection matrix $P$ into the $m$-dimensional cartesian space, usually much lower dimensional, forming the reachable space polytope $\mathcal{P}_x$.
\begin{equation}
    \mathcal{P}_x = \{ \bm{x} \in \mathbf{R}^m \quad| \quad \bm{x}=P\bm{\tau}, \quad \bm{\tau} \in \mathcal{P}_\tau ~\}
\label{eq:polytope_projection}
\end{equation}

The most straight forward way of approaching the enumeration of this polytope is to first enumerate the vertices of the polytope $\mathcal{P}_\tau$, project them to the cartesian space using the projection matrix $P$, followed by a Convex-Hull algorithm to extract the real vertices $\bm{v}_i$ of the polytope $\mathcal{P}_x$ from the projected points and its half-space representation.

In order to avoid enumerating all the vertices of the high-dimensional joint torque polytope $\mathcal{P}_\tau$, as the most complex operation of this polytope enumeration, \textit{Iterative convex hull method}\cite{skuric2022} (ICHM) is used, allowing to directly enumerate low-dimensional cartesian polytope $\mathcal{P}_x$. This iterative method is based on successively applying \textit{Linear programming} (LP) and \textit{Convex-Hull} (CH) algorithms and is able to find both vertex and half-space representation of this polytope at the same time. ICHM algorithm is defined for a family of sets:

\begin{equation}
\{ ~\bm{x}\in R^{m} ~|~ A\bm{x} = B\bm{y},\quad C\bm{y} \leq \bm{d}~\}, \qquad \bm{y} \in R^n,\quad m \leq n
\end{equation}

Therefore the usage of the ICHM algorithm for enumerating the reachable space polytope is rather straight forward, by setting its matrix $A$ to identity $A=I_{m \times m}$, matrix $B$ becomes the projection matrix $P$, the inequality constraint $C\bm{y} \leq \bm{d}$ then becomes the set of the constraints $A\bm{\tau}\leq\bm{b}$.

\section{Analysing the approximation accuracy}
\label{ch:analysis}

Defining metrics of interest for accuracy analysis is highly dependent of the applications these methods will be used on. If the application requires an under-approximation of the reachable set or an over-approximation, the metrics of interest will not be the same. In this paper, the algorithms accuracy is analysed using three different quantitative metrics, each one providing an insight in different characteristics and limitations of the method. 

Additionally as this approximation method has a potential to be used for real-time applications, execution time analysis is performed as well. 

In the extent of the proposed experiments, the considered reachable space is the reachable space of the robot's end effector. 

\subsection{Metrics definition}

\begin{figure}[!t]
    \centering
    \includegraphics[width=\linewidth]{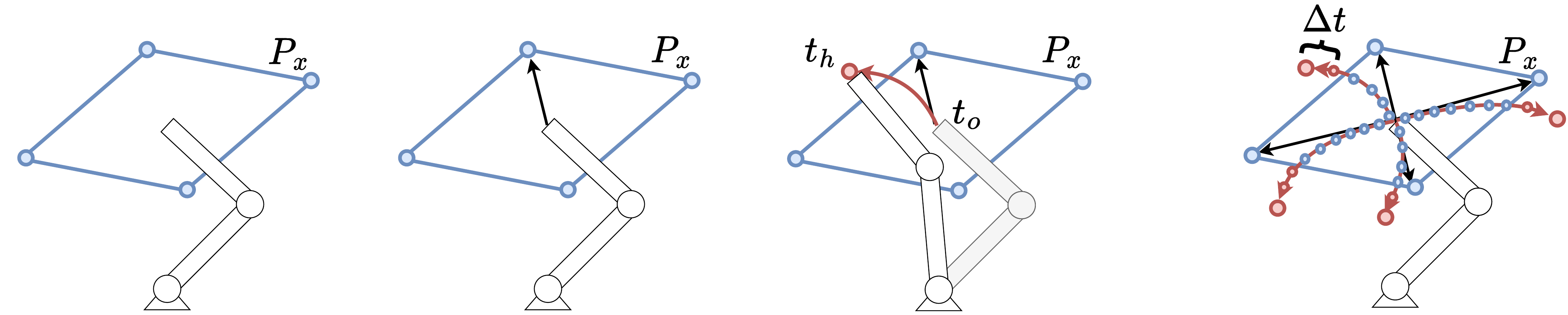}
    \caption{Images show the steps (left to right) of the nonlinear simulation procedure on a simplified 2 link planar robot. First the polytope $\mathcal{P}_x$ of the end-effector is determined. For each vertex of the $\mathcal{P}_x$, joint torque $\tau$ generating the vertex is applied to the robot simulation (figures 2 and 3). All the robot's end-effector positions in all the simulation time steps $\Delta t$ are saved for further analysis.}
    \label{fig:simulation}
\vspace{-0.3cm}
\end{figure}

To analyse the approximation accuracy of the  reachable space polytope $\mathcal{P}_x$, a simple discrete non-linear robot dynamics simulation, subject to the constraints (\ref{eq:limits}) is employed. Simulation is carried out by discretizing the equation (\ref{eq:robot_model}) to determine the instantaneous joint acceleration $\ddot{\bm{q}}_{k}$ and followed by its numerical integration to determine joint velocity $\dot{\bm{q}}_{k+1}$ and position $\bm{q}_{k+1}$ in the step $k\!+\!1$. 
\begin{equation}
\begin{split}
    \ddot{\bm{q}}_{k} &= M^{-1}(\bm{q}_k)\left(\bm{\tau} - C(\bm{q}_k,\dot{\bm{q}}_k)\dot{\bm{q}}_k + \bm{\tau}_g(\bm{q}_k)\right)\\
    \dot{\bm{q}}_{k+1} &= \ddot{\bm{q}}_k\Delta t + \dot{\bm{q}}_{k}\\
    \bm{q}_{k+1} &= \ddot{\bm{q}}_k\frac{\Delta t^2}{2} + \dot{\bm{q}}_k\Delta t  + \bm{q}_{k}\\
\end{split}
\label{eq:simulation_imp}
\end{equation}
Where $\Delta t$ is a fixed simulation sampling time. The cartesian position in each step $\bm{x}_{k}$ is then calculated by evaluating the  robot's forward kinematics
\begin{equation}
    \bm{x}_{k} = f_{DK}(\bm{q}_{k}).
\end{equation}
Each simulation roll-out considers a fixed joint torque vector $\bm{\tau}$ and simulates the robot dynamics through the horizon time $t_h$ performing $N= t_h/\Delta t$ steps.

Each vertex $\bm{v}_i$ of reachable space polytope $\mathcal{P}_x$, calculated for a joint position  $\bm{q}_k$ and a horizon $t_h$, is generated by a joint torque vector $\bm{\tau}_i$, belonging to a subset of vertices of the joint torque polytope $\mathcal{P}_\tau$.

To evaluate the accuracy of the proposed method, the difference in reached space produced by the constant liner model (polytope $\mathcal{P}_x$) and time varying non-linear model of the robot is evaluated, for the same set of applied joint torques. The applied joint torques chosen for the experiments are the joint torques $\bm{\tau}_i$ generating the vertices $\bm{v}_i$ of the reachable space polytope $\mathcal{P}_x$ calculated by the linear model.

As shown on figure  \ref{fig:simulation}, for each of the $n_v$ vertices $\bm{v}_i$ of the polytope $\mathcal{P}_x$ the simulation (\ref{eq:simulation_imp}) is performed and all the cartesian positions $\bm{x}_k$ of the robot in each of the $N=t_h/\Delta t$ sample times are retained for the further analysis. $$
\mathcal{X} = \{\underbrace{\bm{x}_{0,0}~\dots ~\bm{x}_{0,N}}_{\bm{\tau}_0},~\underbrace{\bm{x}_{1,0}~\dots~\bm{x}_{1,N}}_{\bm{\tau}_1},~\dots, ~ \underbrace{\bm{x}_{n_v,0}~\dots~\bm{x}_{n_v,N}}_{\bm{\tau}_{n_v}}\}
$$

\begin{figure}[!t]
    \centering
    \includegraphics[width=0.8\textwidth]{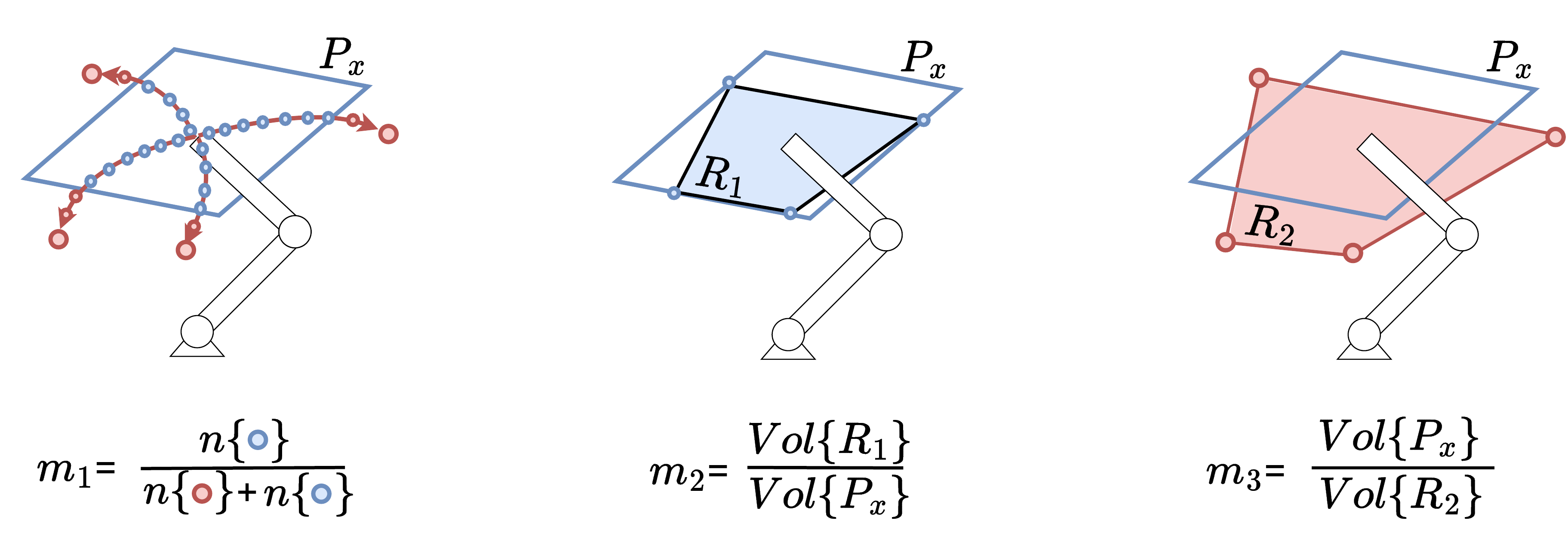}
    \caption{Depiction of the defined quantitative metrics for a simple 2 link planar robot. }
    \label{fig:metrics_defintition}
\vspace{-0.3cm}
\end{figure}

First metric, $m_1$, is defined as a ratio of number of the simulated cartesian robot positions $\bm{x}_k\in \mathcal{X}$ inside the polytope $\mathcal{P}_x$ with respect to the overall number of simulated positions. 

\begin{equation}
    m_1 = \frac{|\mathcal{X}\cap\mathcal{P}_x|}{|\mathcal{X}|} = \frac{|\mathcal{X}\cap\mathcal{P}_x|}{N\cdot n_v}
\end{equation}
The metric $m_1$ gives an insight on how well the polytope $\mathcal{P}_x$ encapsulates the real reachable space determined with simulations. This metric might be used to asses the confidence in the polytope $\mathcal{P}_x$ for safety applications, the closer this metric is to 1, the less chance that the robot can exit the space bounded by $\mathcal{P}_x$.

The second metric, $m_2$, is defined as the ratio of the volumes of the polytope $\mathcal{P}_x$ and the convex hull $CH$ of the points $\bm{x}_k\in\mathcal{X}\cap\mathcal{P}_x$, $$ R_1 = CH(\mathcal{X}\cap\mathcal{P}_x)$$ representing a subset of the points reached by the robot in the simulations found inside of the polytope $\mathcal{P}_x$ .

\begin{equation}
    m_2 = \frac{Vol\{ R_1 \}}{Vol\{P_x\}}
\end{equation}
The metric $m_2$ gives an insight on how much volume of the polytope $\mathcal{P}_x$ is actually attainable by the robot, with respect to the simulations. This metric is designed to asses the quality of the approximation, if the score on this metric is low, this means that the large part of the $\mathcal{P}_x$ is not actually reachable.

The third metric, $m_3$, is defined as the ratio of the volumes of the polytope $\mathcal{P}_x$ and the convex hull $CH$ of all the points $\bm{x}_k\in\mathcal{X}$, $$ R_2 = CH(\mathcal{X})$$ reached by the robot in the simulations.

\begin{equation}
    m_3 = \frac{Vol\{P_x\}}{Vol\{ R_2 \}}
\end{equation}
The metric $m_3$ gives insight about the ratio of the volumes of the polytope $\mathcal{P}_x$ and the simulated reachable space of the robot $CH(\mathcal{X})$. This metric is designed to asses the quality of the volume of the $\mathcal{P}_x$ as a metric. The further this metric from 1 the less confidence one has in the volume of the $\mathcal{P}_x$.

\subsection{Benchmark comparison}
\vspace{-0.1cm}
To benchmark the accuracy of convex polytope approach $\mathcal{P}_x$, it is compared with the coarse approximation of the robot's end-effector reachable space using fixed cartesian velocity $\dot{\bm{x}}$ and acceleration $\ddot{\bm{x}}$ limits specified by robot manufacturer.
\begin{equation}
    \ddot{\bm{x}}\in [~ \ddot{\bm{x}}_{min},  \ddot{\bm{x}}_{max}], \qquad \dot{\bm{x}} \in [~ \dot{\bm{x}}_{min},  \dot{\bm{x}}_{max}]
    \label{eq:limits_cart}
\end{equation}

\begin{figure}[!t]
    \centering
    \includegraphics[width=\linewidth]{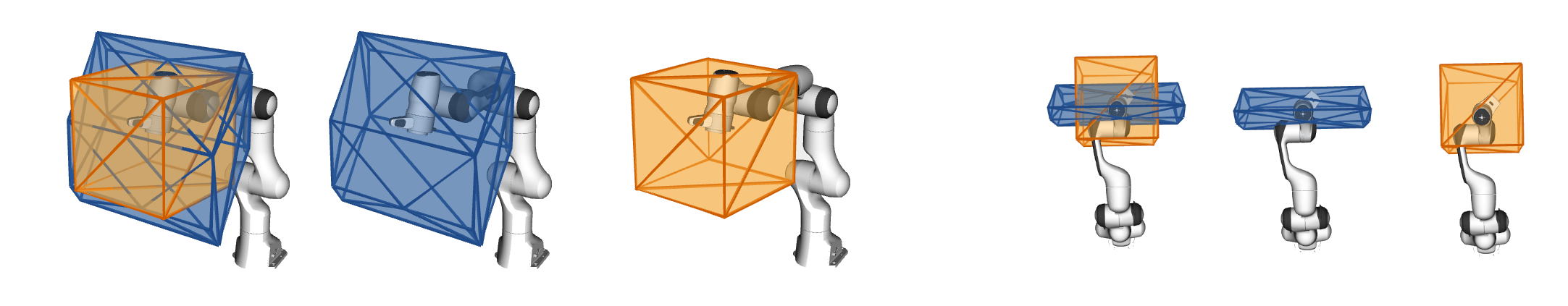}
    \caption{Comparison of the end-effector reachable space approximation using the convex polytope $\mathcal{P}_x$ (blue) and the cube $\mathcal{C}_x$ (orange) for the \textit{Franka Emika Panda} robot with the horizon time $t_h$=200ms. They are both evaluated in robot's initial position and the figures show two different views.}
    \label{fig:cube}
\vspace{-0.3cm}
\end{figure}

The cartesian space limitations (\ref{eq:limits_cart}) assume that the robot's acceleration $\ddot{\bm{x}}$ and velocity $\dot{\bm{x}}$ capacity is constant, which is a not true since the robot's capacity depends highly on robot's joint configuration $\bm{q}$ and velocity  $\dot{\bm{q}}$ \cite{Bowling2005}\cite{skuric2021}. However, they are very convenient and widely used in many applications. A simple convex space $\mathcal{C}_x$ (cube in 3d) of reachable end-effector positions given the cartesian space limitations (\ref{eq:limits_cart}) can be calculated
\begin{equation}
\begin{split}
    \mathcal{C}_x = \{ ~\bm{x}_{k+1} \in R^m ~|~ \bm{x}_{k+1} &= \ddot{\bm{x}}_{k}\frac{t_h^2}{2} + \dot{\bm{x}}_kt_h + \bm{x}_k \\
    \ddot{\bm{x}}_{k} &\in [~ \ddot{\bm{x}}_{min},  \ddot{\bm{x}}_{max}]\\
    \ddot{\bm{x}}_{k}t_h &\in [~ \dot{\bm{x}}_{min},  \dot{\bm{x}}_{max}] ~\}
\end{split}
\end{equation}

Figure \ref{fig:cube} shows the visual comparison of the two approaches: $\mathcal{P}_x$ and $\mathcal{C}_x$ calculated for one pose of the \textit{Franka Emika Panda} robot. 

\vspace{-0.2cm}
\section{Results}
\label{ch:results}
\vspace{-0.2cm}
The proposed analysis of the reachable space polytope accuracy is illustrated on the simulation of the collaborative robotic manipulator \textit{Franka Emika Panda} with 7 degrees of freedom.
Robot modeling and kinematics as well as simulations of robot dynamics are built using the python implementation of the \textsc{robotics-toolbox}\cite{rtb} package. 

The evaluation of the reachable space polytope $\mathcal{P}_x$ and $\mathcal{C}_x$ is carried out using the ICHM algorithm's efficient python implementation within the \textsc{pycapacity}\cite{pycapacity} package. The ICHM's approximation accuracy used for all the evaluations is $\delta$=1mm.

Eight different horizon lengths $t_h$ are chosen for the experiments
$$
t_h = [0.05s, ~ 0.15s, ~ 0.25s, ~ 0.5s,~ 0.75s,~ 1.0s,~ 1.5s, ~ 2.0s]
$$
and the simulation sample time has been set to $\Delta t$=5ms. 

The proposed method implementation as well as the code used for the experiments is fully open-source and can be found in the gitlab repository\footnote{\url{https://gitlab.inria.fr/auctus-team/people/antunskuric/reachable_space/}}. The gitlab repository additionally contains a Robot operating system (ROS) implementation of the proposed method for more interactive evaluation.

All the simulations are run on a computer with 1.90GHz Intel i7-8650U processor and 32Gb of RAM memory.
 
\vspace{-0.1cm}
\subsection{Execution time}
\vspace{-0.1cm}
\begin{figure}[!t]
    \centering
    \includegraphics[width=0.7\linewidth]{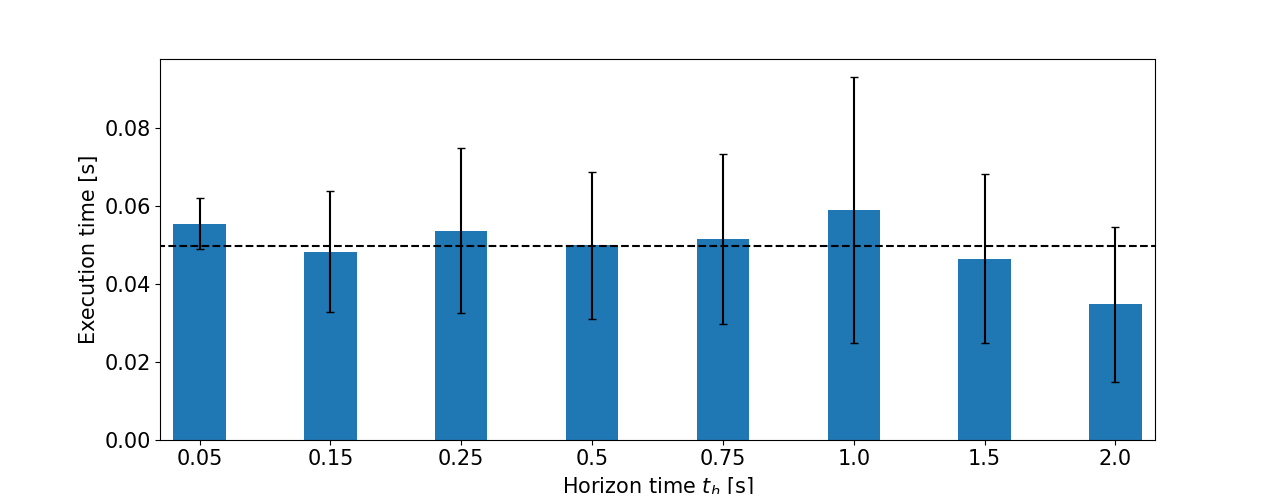}
    \vspace{-0.2cm}
    \caption{Execution time of the polytope $\mathcal{P}_x$ enumeration using ICHM algorithm. Average and standard deviation calculated over 1000 executions for each of the 8 horizon lengths.}
    \label{fig:exec_time}
\vspace{-0.3cm}
\end{figure}

Figure \ref{fig:exec_time} shows the polytope $\mathcal{P}_x$ execution time averaged over 1000 random robot configurations for each of the 8 horizon lengths. From the figure it can be seen that the execution time is relatively consistent through all the horizon lengths, with the average value of around 50ms. The constant execution time is expected because the horizon length $t_h$ does not have a big influence on the geometrical complexity of the polytope (number of vertices and faces) but its scale, and the scale itself does have a big influence the execution time.

\begin{figure}[!t]
    \centering
    \includegraphics[width=0.75\linewidth]{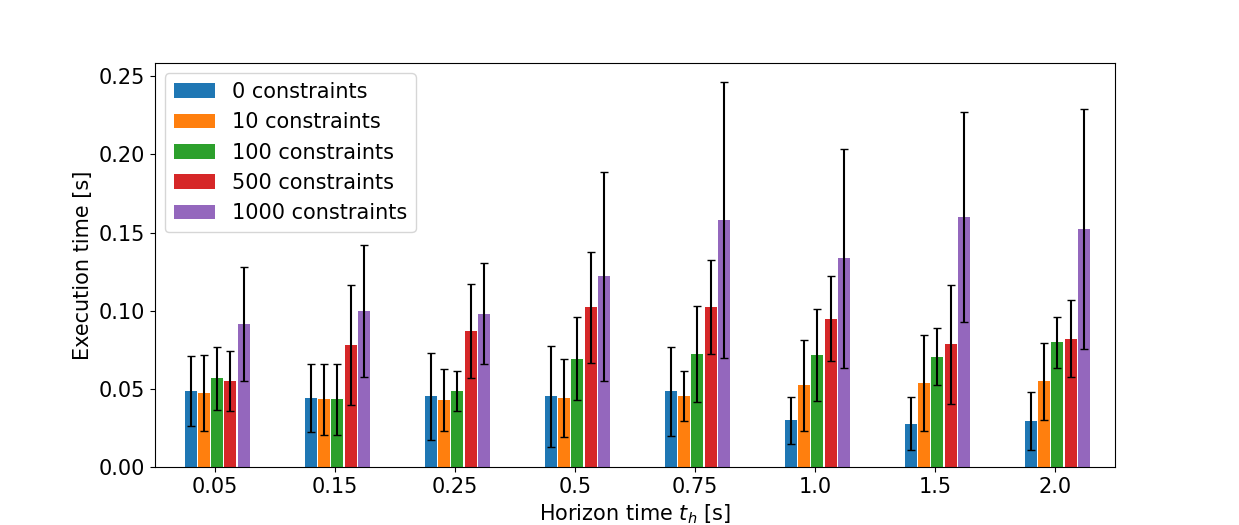}
    \caption{Average execution time and standard deviation calculated over 1000 executions for each of the 8 horizon lengths, with different number of added environmental constraints, ranging from 0 to 1000.}
    \label{fig:exec_time_constraints}
    \vspace{-0.3cm}
\end{figure}

Figure \ref{fig:exec_time_constraints} shows the ICHM execution time for the 8 horizon lengths with different numbers of environmental constraints $A_e\bm{x}\!\leq\!\bm{b}_e$. All the environmental constraints are generated randomly and all the results are averaged over 1000 random robot configurations for each horizon length $t_h$ and all 4 different numbers of environmental constraints: 0 (no environment constraints), 10, 100, 500 and 1000. Results show that up to 10 added environmental constraints the execution time of the method does not change significantly, having the average execution time around 50ms. With 100 environmental constraints the average execution time increases around 20\% to around 70ms, for 500 the average execution is 100ms and with 1000 constraints the average execution time more than doubles, to 150ms.

The execution time analysis of the ICHM algorithm has shown that the average polytope $\mathcal{P}_x$ enumeration takes between 50 and 150ms, which opens many doors for potential real-time applications.

The average execution time of the nonlinear simulation for $t_h$=50ms and $\Delta t$=5ms is around 60s, and it grows to about 240s for $t_h$=2s. The cube $\mathcal{C}_x$ average execution time stays constant for all horizon lengths at around 2ms.

\vspace{-0.3cm}
\subsection{Accuracy analysis}
\begin{figure}[!t]
    \centering
    \includegraphics[width=0.95\textwidth]{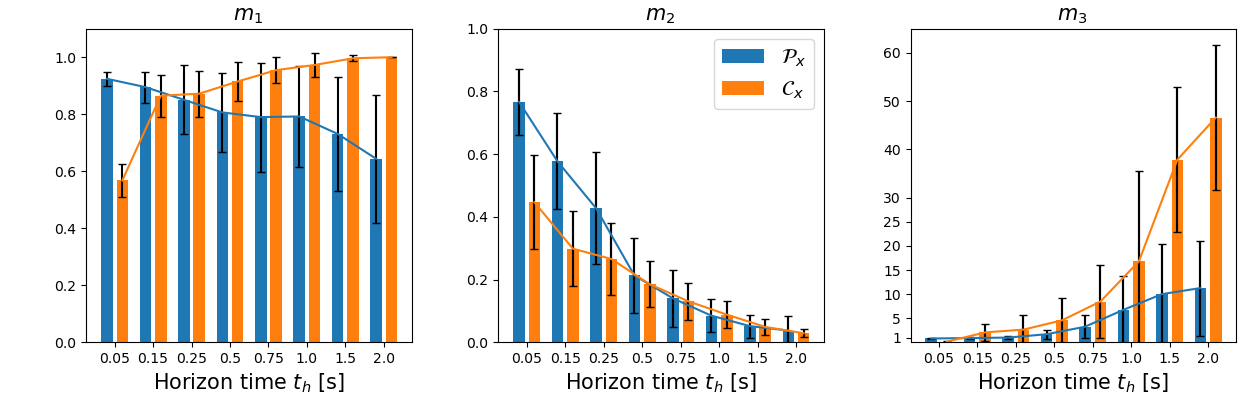}
    \caption{Accuracy measures calculated for $\mathcal{P}_x$ and $\mathcal{C}_x$, averaged over 100 random poses of the \textit{Franka Emika panda} robot for each of 8 horizon lengths.}
    \label{fig:accuracy}
    \vspace{-0.5cm}
\end{figure}

\vspace{-0.1cm}
Figure \ref{fig:accuracy} shows the evolution of the three metrics $m_1$, $m_2$ and $m_3$ with respect to the horizon length $t_h$ for the $\mathcal{P}_x$ and $\mathcal{C}_x$, averaged over 100 random robot configurations. 

The first graph on the figure \ref{fig:accuracy} shows the metric $m_1$. It can be seen that the ratio of simulated robot positions $\bm{x}_k \in \mathcal{X}$ inside the polytope $\mathcal{P}_x$ decreases with the increasing horizon length $t_h$, while it increases for the $\mathcal{C}_x$. For larger horizon lengths the robot is able to move far from the initial joint configuration $\bm{q}$ at the beginning of the horizon, making the assumption of linearity of the robot model, made for calculating $\mathcal{P}_x$, largely inaccurate. The approach $\mathcal{C}_x$, on the other hand, improves with the increase in horizon length, since the space bounded by the cube $\mathcal{C}_x$ increases linearly with the horizon length $t_h$, at some point it encapsulates the whole workspace of the robot. However, it can be seen that even though the ratio $m_1$ decreases, the polytope $\mathcal{P}_x$ still contains the most ($\geq$60\%) of the simulated robot's reached positions $\bm{x}_k\in\mathcal{X}$. Finally, it can also be seen that the variance of the $m_1$ increases considerably, lowering the confidence of the polytope $\mathcal{P}_x$ for longer horizons.

Evolution of the metric $m_2$ is shown on the middle graph of the figure \ref{fig:accuracy} . It can be seen that the volume of all the reached robot positions inside polytope $\mathcal{P}_x$ and $\mathcal{C}_x$ decreases considerably with the increase of the horizon time $t_h$. For the horizon times under $t_h\leq 0.25$s the majority ($\geq$50\%) of the reachable space polytope $\mathcal{P}_x$ has been reached by the robot simulations. However for the longer horizon times $t_h\geq0.5$s the reached volume ratio $m_2$ drops under 20\% of the total volume of the $\mathcal{P}_x$. In the case of $\mathcal{C}_x$, the majority ($\geq$50\%)  of the volume of the $\mathcal{C}_x$ is never reached by the robot for all tested horizon lengths $t_h$. 


The right graph of the figure \ref{fig:accuracy} shows the evolution of the metric $m_3$. The graph shows that the volume of the polytope $\mathcal{P}_x$ and the cube $\mathcal{C}_x$ becomes several times higher than the volume of the $R_2$ with the increase in the horizon time $t_h$. For the horizon time of $t_h$=2s the average volume of $\mathcal{P}_x$ is 12 times larger than the volume of $R_2$ and the volume $\mathcal{C}_x$ is more than 50 times larger than $R_2$. At the same time the average reached volume $R_1$ inside the $\mathcal{P}_x$ and $\mathcal{C}_x$ consists of under 10\% of their volume (metric $m_2$), which potentially makes both polytope $\mathcal{P}_x$ and $\mathcal{C}_x$ impractical for many of applications. 

However, for the shorter horizon times $t_h\leq0.25$s the ratio $m_3$ is very close to 1 for $\mathcal{P}_x$, which means that the reachable space polytope's $\mathcal{P}_x$ volume corresponds well to the actual reached volume of the robot $R_2$.

Overall, this numerical analysis shows that the reachable space approximation based on the convex polytope $\mathcal{P}_x$ works reasonably well for low horizon times $t_h\leq250$ms, while for the longer horizon times $t_h>250$ms the assumption of the linearity of the system produces a large error in the reachable space estimation, producing the polytope $\mathcal{P}_x$ with volume several times higher than the real reachable space of the robot (shown by the metric $m_3$), most of which is not reachable by the robot at all (showed by the metric $m_2$). Finally, the results show that the polytope $\mathcal{P}_x$, even though a coarse approximation, has better performance across all the defined metrics then the $\mathcal{C}_x$ based on the constant cartesian limits, at the same time having smaller average volume for the all the horizon times. These results remind, once again, that robot's cartesian capacity is not constant and by modeling their evolution, even in a coarse manner, a substantial improvement in the estimation accuracy can be reached.
\vspace{-0.3cm}
\section{Discussion}
\label{ch:discussion}
\vspace{-0.2cm}
Approximating the reachable space of the robot using convex polytopes $\mathcal{P}_x$ is an efficient and relatively fast way of gaining information about the potential robot's position for the horizon of interest. Even though this approximation approach is neither an under or an over approximation of the true reachable space of the robot, as shown in the numerical analysis, for shorter horizon times $t_h\leq250$ms this metric is reasonably precise. Due to its efficiency, easy and intuitive integration of different environmental constraints, possibility to consider the robot's link geometry and the fact that its output has the form of a convex set of inequality constraints (convex polytope $\mathcal{P}_x$), it has a potential to be a useful tool for robot performance and safety analysis. However there are several important limitations of this approach.

\subsection{Main limitations}
First of all, the reachable space of a robot is highly non-convex and non-linear. As shown in the numerical analysis section \ref{ch:results}, approximating this space with convex polytopes is relatively precise only for small horizon times $t_h\leq250$ms.

Secondly, robot dynamics is highly non-linear and time variant as well. When calculating the reachable space polytope $\mathcal{P}_x$, as formulated in section \ref{ch:polytope}, the robot model is linearised in the robot's joint configuration $\bm{q}_k$ and velocity $\dot{\bm{q}}_k$  at the beginning of the horizon and this linearised model is considered constant during the horizon length. The consideration of constant robot model and its linearity, is only valid for small robot displacements around the current robot 
state, which is again a strong assumption and only valid for short horizon times. 

Finally, the polytope formulation from the section \ref{ch:polytope}, considers only constant joint torques $\bm{\tau}$ applied to the robot during the whole horizon length $t_h$. The polytope $\mathcal{P}_x$ calculates the joint torque vectors $\bm{\tau}$ that will produce the largest possible displacements in cartesian space given the robot and environment constraints. This calculation is based on the robot's linearised model at the beginning of the horizon, which in term, once again, limits the precision of the approximation to the small displacements around the initial robot state, preventing longer horizon lengths $t_h$, as confirmed in the analysis section \ref{ch:results}.

\vspace{-0.3cm}

\subsection{Potential applications}

One promising field of applications for the polytope based reachable space metric is the human-robot interaction domain. The polytope is essentially a triangulated mesh and can be easily visualised, which could be an easy and intuitive way to give insight to the human operators about robot's real-time capabilities. It has a potential to be used for the haptic control applications, as human operators might not be always aware of how close the robot is to the limitation of its capabilities or what is the possible space the robot can be in while teleoperating. Furthermore, this metric could be potentially used to to increase human operators awareness about the robot's state during the co-manipulation by visualising the polytope $\mathcal{P}_x$ in real-time and even potentially increase the interaction safety. 

The numerical analysis has shown that for shorter horizon times $t_h\leq250$ms the volume of the polytope $\mathcal{P}_x$ is comparable to the volume of the simulated reached space $R_2$, as shown on the figure \ref{fig:accuracy}. The volume of the $\mathcal{P}_x$ could potentially be used as a performance metric for robotic workspace analysis and for applications such as robot design.

Finally, as the polytope $\mathcal{P}_x$ can be represented as a set of convex inequalities $H\bm{x}\leq\bm{d}$, this metric could potentially be used as a set of constraints for robot control. For example in case of the Model Predictive Control (MPC) where, in many cases, robot's dynamics is already considered linear and constant during given horizon time, which corresponds well with the assumptions of the polytope $\mathcal{P}_x$ formulation. 



\section{Conclusion}
In this paper a convex polytope based robot's reachable space approximation metric is introduced. The metric predicts the robot's reachable space for a given time horizon based on the robot's actuation capabilities and its kinematic constraints. The proposed metric can intuitively integrate the environmental constraints as well as the robot's link geometry. With the execution time of around 50ms, for standard 7dof collaborative robot, it has a potential to be used for real-time applications. 

However, the metric relies on a linearised robot model, which limits its use to shorter horizon times. The analysis has shown that the accuracy of the approximation decreases significantly for horizon times longer than 250ms. 
\vspace{-0.2cm}




%
%
%
\bibliography{hfr2022.bib}
\bibliographystyle{splncs04}




\end{document}